\documentclass{ws-procs975x65}

\begin{document}

\title{CONFORMAL FORMULATION OF COSMOLOGICAL FUTURES}

\author{PHILIPP A. H\"OHN$^{1,2}$ and SUSAN M. SCOTT$^2$}

\address{$^1$Institute for Theoretical Physics, Universiteit Utrecht,\\
  Leuvenlaan 4, NL-3584 CE Utrecht, The Netherlands.\\
  $^2$Centre for Gravitational Physics, Department of Quantum Science, College of Physical Sciences, The Australian National University, Canberra ACT 0200, Australia.\\
  E-mail: p.a.hohn@uu.nl, susan.scott@anu.edu.au}

\begin{abstract}
We summarise the new conformal framework of an \emph{anisotropic future endless universe} and an \emph{anisotropic future singularity}. Both new definitions are motivated by, but not restricted to quiescent cosmology and the Weyl curvature hypothesis, which previously only possessed a framework for a classical initial state of the universe, namely the isotropic singularity. Some of the features of the framework are briefly discussed.
\end{abstract}

\bigskip
\bodymatter
\bigskip
In a first attempt to explain the apparent isotropy in the observable vicinity of our universe, Misner\cite{Misner} introduced his idea of \emph{chaotic cosmology} according to which the infant universe must have been highly chaotic and irregular before dissipative effects eventually ironed out the geometry and matter distribution. The appeal of this idea lies in the conclusion that the current state of the universe would be largely independent of the initial state; in a sense our current universe would be ``logically inevitable''. There have been indications, however, that such a chaotic explanation is not viable in its full generality\cite{Hawking, Penrose1,Penrose2, Barrow}.

As a direct competitor, the idea of \emph{quiescent cosmology} was introduced by Barrow\cite{Barrow}, claiming that the universe originated in an already smooth and regular initial state with matter being in thermal equilibrium. A possible explanation for the necessary rise of entropy over the apparent high initial value was provided by Penrose who argued for a connection between the contribution of the gravitational field to the total entropy, gravitational clumping and the Weyl curvature, which culminated in the so-called Weyl curvature hypothesis (WCH). According to this conjecture, the Weyl curvature was initially bounded by the matter controlled Ricci curvature while the opposite should be true at a cosmological future\cite{Penrose1,Penrose2}. In fact, the WCH is, moreover, not incompatible with the inflationary paradigm which lacks an explanation for the regularity of the patch which is inflated into our observable universe. 

In the light of \emph{quiescent cosmology} and the WCH, a classical initial state of the universe should thus be given by an isotropic Friedman-type singularity while a cosmological future should involve irregularities and a high degree of gravitational clumping. A mathematical framework, incorporating the ideas of this school of thought, already exists for the initial state, namely the definition of an \emph{isotropic past singularity} (IPS)\cite{GW}, whose essential ingredients can be summarised as follows:
\begin{definition}[Isotropic Past Singularity (IPS)]
The physical spacetime $(\mathcal{M},\mathbf{g})$ is said to admit an IPS if it can be related to an unphysical spacetime $(\tilde{\mathcal{M}},\mathbf{\tilde{g}})$ via a conformal rescaling $\mathbf{g}=\Omega^2(T)\mathbf{\tilde{g}}$, satisfying the following conditions:
\begin{enumerate}
\item $T$ is a smooth cosmic time function on $(\tilde{\mathcal{M}},\mathbf{\tilde{g}})$ and $\mathcal{M}$ is the open submanifold $T>0$,
\item a regularity condition for $\mathbf{\tilde{g}}$ on the slice $T=0$,
\item $\Omega(0)=0$, and
\item some general conditions on the conformal factor $\Omega(T)$.
\end{enumerate}
\end{definition}
While the regularity condition on $\mathbf{\tilde{g}}$ on the slice $T=0$ ensures that this slice is entirely smooth and regular in the unphysical conformal spacetime $(\tilde{\mathcal{M}},\mathbf{\tilde{g}})$, the vanishing of the conformal factor at $T=0$ renders the physical spacetime $(\mathcal{M},\mathbf{g})$ singular with diverging curvature invariants. It can also be shown that this definition--together with conformal regularity conditions on the cosmological fluid flow $\mathbf{u}$--guarantees an asymptotic initial isotropy and initial compatibility with the WCH\cite{GW}, since at the IPS
\begin{eqnarray}
\mathop {\lim }\limits_{T \to 0^{+}}\frac{C_{abcd}C^{abcd}}{R_{ef}R^{ef}}=\mathop {\lim }\limits_{T \to
0^{+}}\frac{\sigma^{2}}{\theta^{2}}=\mathop {\lim
}\limits_{T \to 0^{+}}\frac{\omega^{2}}{\theta^{2}}=\mathop {\lim }\limits_{T \to
0^{+}} \frac{\dot{u}^{a}\dot{u}_{a}}{\theta^{2}}=0 \qquad,
\label{kiniso}
\end{eqnarray}
where $\sigma, \omega$ and $\dot{u}$ are the shear, vorticity and acceleration, i.e., the anisotropic kinematical quantities, of the flow $\mathbf{u}$, respectively, while $\theta$ denotes the isotropic expansion scalar. In consequence of the above, we refer to the slice $T=0$ as the IPS. The regularity in $(\tilde{\mathcal{M}},\mathbf{\tilde{g}})$ renders this definition analytically especially advantageous. 

Projecting the WCH forward in time, we expect the high entropy state of a cosmological future to be associated with a high degree of (local) gravitational clumping which should (at least locally) lead to the development of anisotropies with cosmic evolution. The definition of the IPS, however, admits numerous FRW example cosmologies which implies that this framework is not alone sufficient to guarantee an irregular future evolution compatible with the WCH. 

With the intent of devising an analogous conformal framework for cosmological futures which admits anisotropies, one can show that under fairly general conditions a regular conformal structure with a conformal factor which is merely a function of cosmic time $T$ necessarily leads to an asymptotic isotropy in the sense of (\ref{kiniso}), irrespective of whether $T$ approaches zero from below (corresponding to a cosmological future) or above (corresponding to an initial state) or whether the conformal factor vanishes or diverges at $T=0$\cite{hoehnscott1}. This result highlights the special geometric nature of the IPS-framework, implying that we cannot maintain a high degree of regularity in the conformal structure if we are interested in less symmetric asymptotic scenarios. As a consequence, we expect irregularities in a conformal structure for anisotropic cosmological futures constructed along the lines above. 

In example cosmologies which admit an IPS and which, furthermore, possess cosmological futures compatible with the WCH, one, indeed, finds irregularities in a conformal rescaling whose conformal factor $\bar{\Omega}(T)$ again is merely a function of cosmic time, where $T=0$ at the cosmological future. These irregularities manifest themselves in the covariant conformal metric becoming degenerate as $T\rightarrow 0^{-}$\cite{hoehnscott1,hoehnscott2}.

Motivated by general results, as well as example cosmologies, one can then define the irregular conformal framework of an \emph{anisotropic future singularity} (AFS) and--for the ever expanding scenario--of an \emph{anisotropic future endless universe} (AFEU)\cite{hoehnscott1,hoehnscottGRF,hoehnscott2}. We summarise here only the conformal definition of the AFS:
\begin{definition}[Anisotropic Future Singularity (AFS)]
The physical spacetime $(\mathcal{M},\mathbf{g})$ is said to admit an AFS if 
\begin{enumerate}
\item $\exists$ a smooth function $\bar{T}$ on $\bar{\mathcal{M}}\supset\mathcal{M}$ such that $\mathcal{M}$ is the open submanifold $\bar{T}<0$,
\item $\exists$ a continuous tensor field $\mathbf{\bar{g}}$ of type $(0,2)$ on $\mathcal{M}\cup\mathcal{N}$, where $\mathcal{N}$ is an open neighbourhood of $\bar{T}=0$,
\item $\bar{T}$ is a smooth cosmic time function with respect to $\mathbf{\bar{g}}$ on $\mathcal{M}\cup\mathcal{N}$,
\item $\mathbf{g}=\bar{\Omega}^2(\bar{T})\mathbf{\bar{g}}$ on $\mathcal{M}$ and $\mathbf{\bar{g}}$ is degenerate on $\bar{T}=0$,
\item $\mathop {\lim }\limits_{\bar{T} \to 0^{-}}\bar{\Omega}(\bar{T})=+\infty$, $\mathop {\lim }\limits_{\bar{T} \to 0^{-}}\det\left(\bar{\Omega}^2(\bar{T})\mathbf{\bar{g}}\right)=0$ and some general conditions on $\bar{\Omega}(\bar{T})$.
\end{enumerate}
\end{definition}
The degeneracy causes a singularity for both the physical and conformal spacetime. Nevertheless, due to the continuity of $\mathbf{\bar{g}}$ and the fact that the singular part has been absorbed in $\bar{\Omega}$, the conformal spacetime is more regular, which is important from an analytical point of view and does not render the above framework dysfunctional. 

As example cosmologies of the AFS/AFEU-framework show, the new definitions admit a great variety of anisotropic future evolution\cite{hoehnscott1,hoehnscott2}. Furthermore, in accordance with the recollapsing and ever-expanding scenarios, respectively, it can be shown that $\theta$ is negative prior to reaching the AFS and positive prior to $\bar{T}=0$ in an AFEU. Under relatively general conditions, it can also be proved that the AFS is a \emph{deformationally strong singularity}, thereby emphasising the final cosmological state\cite{hoehnscott2}. Finally, the conformal framework employed in the new definitions is, interestingly, somewhat dual to the conformal framework used in the \emph{dynamical systems} approach to cosmology\cite{hoehnscott1} which has been in vogue during recent years\cite{dynsys}.

The precise implications of the AFS/AFEU-framework on the degree of anisotropy remain to be derived. We conjecture, nonetheless, that the conjunction of the IPS with the new conformal framework provides a possible version of a complete formalisation of \emph{quiescent cosmology} and the WCH.


\begin{thebibliography}{9}
\bibitem{Misner} C.~W.~Misner, {\em ApJ} {\bf 151}, 431 (1968)
\bibitem{Hawking} C.~B.~Collins and S.~W.~Hawking, {\em ApJ} {\bf 180}, 317 (1973)
\bibitem{Penrose1} R.~Penrose, in {\it General Relativity: An Einstein Centenary Survey}, ed S.~W.~Hawking and W.~Israel (Cambridge: Cambridge University Press, 1979)
\bibitem{Penrose2} R.~Penrose, {\em Ann.\ N.\ Y.\ Acad.\ Sci. } {\bf 571 (1)}, 249 (1989)
\bibitem{Barrow} J.~D.~Barrow, {\em Nature} {\bf 272}, 211 (1978)
\bibitem{GW} S.~W.~Goode and J.~Wainwright, {\em Class.\ Quant.\ Grav.} {\bf 2}, 99 (1985)
\bibitem{hoehnscott1} P.~A.~H\"ohn and S.~M.~Scott, {\em Class.\ Quant.\ Grav.} {\bf 26}, 035019 (2009) [arXiv:1001.2928]
\bibitem{hoehnscottGRF} P.~A.~H\"ohn and S.~M.~Scott, {\em Int.\ J.\ Mod.\ Phys.\ D} {\bf 17}, 571 (2008) [arXiv:1001.2823]
\bibitem{hoehnscott2} P.~A.~H\"ohn and S.~M.~Scott, {\it in preparation}
\bibitem{dynsys} G.~F.~R.~Ellis and J.~Wainwright (eds) {\it Dynamical Systems in Cosmology} (Cambridge: Cambridge University Press, 1997)
\end{thebibliography}
\end{document}